\documentclass[aps,prd,preprint,superscriptaddress,amsmath,amssymb,showpacs]{revtex4-1}
\usepackage{dcolumn}
\usepackage{graphicx}
\usepackage{float}
\usepackage{physics}
\usepackage[colorlinks=true,allcolors=blue]{hyperref}
\begin{document}
	\title{Causality and Stability of First-Order Relativistic Spin Hydrodynamics with Conserved Charges}
	
	\author{Wei Lu}
	\affiliation{College of Mathematics and Physics, China Three Gorges University, Yichang 443002, China}
	
	\author{Yang Zhong}
	\email{zhy@ctgu.edu.cn}
	\affiliation{College of Mathematics and Physics, China Three Gorges University, Yichang 443002, China}
	\affiliation{Center for Astronomy and Space Sciences and Institute of Modern Physics, China Three Gorges University, Yichang 443002, China}
	
	\author{Sheng-Qin Feng}
	\email{fengsq@ctgu.edu.cn}
	\affiliation{College of Mathematics and Physics, China Three Gorges University, Yichang 443002, China}
	\affiliation{Center for Astronomy and Space Sciences and Institute of Modern Physics, China Three Gorges University, Yichang 443002, China}
	\affiliation{Key Laboratory of Quark and Lepton Physics (MOE) and Institute of Particle Physics,\\
		Central China Normal University, Wuhan 430079, China}	

\date{\today}%

	\begin{abstract}
		 We study the causality and stability of first-order relativistic spin hydrodynamics with particle-number conservation. By deriving the complete dispersion relations of linear perturbations around global equilibrium, we find that conserved-charge dynamics modifies the sound sector and introduces additional non-hydrodynamic modes absent in the charge-neutral theory. While the structure of spin relaxation modes remains unchanged, the stability conditions acquire new contributions from charge diffusion and thermodynamic susceptibilities. More importantly, a particle-number-induced mode is shown to violate the causality condition in the short-wavelength limit. We further demonstrate that particle-number conservation does not remove the instability inherent in first-order spin hydrodynamics. These results reveal nontrivial interplay between spin and conserved-charge dynamics and provide important constraints on relativistic spin hydrodynamic theories at finite density.
	\end{abstract}
	
\maketitle
	
\section{Introduction}	
Relativistic heavy-ion collisions provide a unique opportunity to study strongly interacting matter under extreme conditions of temperature and density, namely the quark-gluon plasma (QGP). In non-central collisions, the colliding nuclei carry an enormous initial orbital angular momentum, reaching the order of $10^{7}\hbar$ at RHIC and LHC energies. Through spin-orbit coupling, part of this orbital angular momentum can be converted into the spin polarization of produced particles. Over the past decade, measurements by the STAR and ALICE Collaborations have established the global polarization of $\Lambda$ and $\bar{\Lambda}$ hyperons \cite{STAR:2020xbm,STAR:2017ckg,Niida:2018hfw,STAR:2021beb,STAR:2019erd,Becattini:2017gcx} and revealed spin-alignment signals of vector mesons \cite{ALICE:2019aid,Singha:2022syo,STAR:2022fan,Schilling:1969um,ALICE:2022dyy}. In particular, the observation of global $\Lambda$ polarization perpendicular to the reaction plane indicates that the QGP created in heavy-ion collisions is the most vortical fluid ever observed experimentally \cite{ALICE:2019aid,Becattini:2024uha,STAR:2018gyt,ALICE:2019onw,STAR:2020xbm,ALICE:2021pzu,Shen:2020mgh,Liang:2004xn,Liang:2004ph}.

At the same time, several experimental observations remain difficult to reconcile with existing theoretical descriptions. Measurements of local polarization exhibit qualitative discrepancies with predictions based on thermal-vorticity-induced polarization \cite{Karpenko:2016jyx,Xie:2017upb,Fu:2020oxj,Wu:2019eyi,Becattini:2017gcx}, while the observed spin alignment of vector mesons such as $\phi$ and $K^{*0}$ cannot be satisfactorily explained within conventional frameworks \cite{Schilling:1969um,Sheng:2019kmk,Sheng:2020ghv,Sheng:2022wsy}. These puzzles strongly suggest that the dynamical evolution of spin degrees of freedom must be incorporated into the theoretical description of relativistic fluids in a self-consistent manner.

Motivated by these developments, relativistic spin hydrodynamics has emerged as an important extension of conventional hydrodynamics \cite{Hongo:2021ona,Wang:2021ngp,Speranza:2021bxf,Li:2020eon,Hattori:2019lfp,Fukushima:2020ucl,She:2021lhe,Florkowski:2018myy,Peng:2021ago,Florkowski:2017ruc,Hidaka:2017auj,Weickgenannt:2019dks,Sheng:2021kfc,Fang:2022ttm,Hu:2021pwh,Weickgenannt:2020aaf,Becattini:2007nd,Becattini:2009wh,Becattini:2012pp,Hu:2021lnx,Becattini:2012tc,Becattini:2018duy,Montenegro:2017rbu,Montenegro:2017lvf,Montenegro:2018bcf,Montenegro:2020paq,Liu:2020dxg,Gallegos:2020otk,Garbiso:2020puw}. In this framework, spin angular momentum is promoted to an independent hydrodynamic degree of freedom whose evolution is coupled to orbital angular momentum through the conservation of total angular momentum. Depending on the choice of energy-momentum tensor and spin tensor, as well as the power-counting assignment of the spin chemical potential $\omega^{\mu\nu}$, several formulations of spin hydrodynamics have been proposed \cite{Hattori:2019lfp,Fukushima:2020ucl,She:2021lhe}. Typical examples include the canonical, GLW, and Belinfante formulations, with $\omega^{\mu\nu}$ treated either as a leading-order quantity, $\mathcal{O}(1)$, or as a first-order gradient correction, $\mathcal{O}(\partial)$. Although these formulations may be related through pseudogauge transformations, they can lead to different local observables, and their phenomenological applicability remains under active investigation \cite{Xie:2023gbo,Daher:2022wzf,Daher:2024bah,Sarwar:2022yzs,Lu:2026ceo}.

Besides phenomenological relevance, any hydrodynamic theory must satisfy the fundamental requirements of causality and stability. Causality demands that no signal propagates faster than light, while stability requires that small perturbations around equilibrium do not grow without bound. For conventional relativistic dissipative hydrodynamics, Hiscock and Lindblom demonstrated that first-order Navier-Stokes theory in the Landau frame is both acausal and unstable \cite{Hiscock:1983zz,Hiscock:1985zz,Hiscock:1987zz}. Similar issues arise in spin hydrodynamics. Recent studies have shown that first-order spin hydrodynamic theories generally suffer from pathological modes, including divergent propagation velocities in the short-wavelength limit and exponentially growing perturbations around equilibrium \cite{Xie:2023gbo,Daher:2022wzf,Daher:2024bah,Sarwar:2022yzs,Lu:2026ceo}. Furthermore, the resulting stability conditions depend sensitively on the spin equation of state and the corresponding thermodynamic susceptibilities.

However, existing analyses have almost exclusively focused on systems without conserved-charge dynamics~\cite{Xie:2023gbo,Daher:2022wzf,Daher:2024bah,Sarwar:2022yzs,Lu:2026ceo}, namely $n=0$ and  $n^\mu=0$. Under~this assumption, only energy, momentum, and~spin fluctuations contribute to the hydrodynamic spectrum.~Although such an approximation may be adequate near midrapidity at top RHIC and LHC energies, finite-density effects become increasingly important away from this regime, particularly in the RHIC Beam Energy Scan program and in lower-energy heavy-ion experiments such as HADES~\cite{Bzdak:2019pkr,Galatyuk:2014vha}. Theoretical descriptions incorporating conserved-charge dynamics are also relevant to the physics programs of facilities such as FAIR, NICA, and~HIAF, which aim to investigate strongly interacting matter at high baryon density and related regions of the QCD phase diagram~\cite{CBM:2016kpk,MPD:2022qhn,Zhou:2022pxl}.

The inclusion of particle-number conservation is not assumed, by~itself, to~cure the known acausality or instability of first-order spin hydrodynamics. Rather, it enlarges the hydrodynamic fluctuation spectrum and modifies the couplings among the charge, energy, momentum, and~spin sectors. First, fluctuations of the conserved current, $N^\mu=nu^\mu+n^\mu$, introduce additional degrees of freedom and couple charge fluctuations to energy, momentum, and~spin fluctuations. As~a result, additional diffusive, propagating, or~non-hydrodynamic modes may emerge, altering the structure of the dispersion relations. Second, the~thermodynamic description is enlarged by the inclusion of the particle-number density and chemical potential, leading to a broader parameter space of susceptibilities and transport coefficients. The~stability and causality conditions obtained in the charge-neutral limit must therefore be reexamined in the presence of conserved-charge~dynamics.

Despite its phenomenological relevance, the~causality and stability of first-order spin hydrodynamics with conserved charges have received relatively limited systematic investigation. It therefore remains unclear how conserved-charge dynamics modifies the pathological modes already present in the charge-neutral theory and whether it generates additional acausal or unstable structures. Addressing these questions constitutes the primary motivation of the present~work.

In this paper, we investigate the causality and stability properties of first-order relativistic spin hydrodynamics in the presence of particle-number conservation. We adopt the canonical formulation and work within the counting scheme $\omega^{\mu\nu}\sim\mathcal{O}(\partial)$, where the spin chemical potential is treated as a first-order quantity in the gradient expansion. The analysis is performed in the Landau frame.

We consider small perturbations around a homogeneous equilibrium state, including fluctuations of the energy density $\delta e$, fluid velocity $\delta u^i$, spin density components $\delta S^{0i}$ and $\delta S^{ij}$, particle-number density $\delta n$, and diffusion current $\delta n^i$. By deriving the complete set of linearized equations and performing a Fourier-mode analysis, we obtain the full dispersion relations of charge, sound, shear, and spin modes. The corresponding causality and stability conditions are then analyzed in both the long-wavelength $(k\rightarrow0)$ and short-wavelength $(k\rightarrow\infty)$ limits. Particular attention is devoted to determining how conserved-charge dynamics modifies the mode spectrum and whether it introduces additional acausal or unstable structures beyond those already present in the charge-neutral theory. We further examine the constraints imposed by thermodynamic susceptibilities and transport coefficients on the physical consistency of the theory.

This paper is organized as follows. In Sec. \ref{chap:2}, we briefly review the fundamental equations and constitutive relations of spin hydrodynamics with particle-number conservation. In Sec. \ref{chap:3}, we derive the dispersion relations and analyze the corresponding causality and stability conditions. Finally, Sec. \ref{chap:4} contains our conclusions and outlook.

Throughout this work, we adopt the metric convention $g_{\mu\nu}=\operatorname{diag} \left\{ +,-,-,-\right\}$, The four-velocity satisfies the normalization condition $u^{\mu}u_{\mu}=1$, and the projection tensor orthogonal to the four-velocity is defined as $\Delta^{\mu\nu}=g^{\mu\nu}-u^{\mu}u^{\nu}$. The projection of a four-vector onto the subspace orthogonal to the four-velocity is denoted by $A^{\left\langle \mu\right\rangle }=\Delta^{\mu\nu}A_{\nu}$. For an arbitrary rank-two tensor, its symmetric and antisymmetric parts are denoted by $A_{(s)}^{\mu\nu}=A^{(\mu\nu)}=\frac{1}{2}\left(A^{\mu\nu}+A^{\nu\mu}\right)$ and $A_{(a)}^{\mu\nu}=A^{[\mu\nu]}=\frac{1}{2}\left(A^{\mu\nu}-A^{\nu\mu}\right)$ respectively. The symmetric, traceless, and four-velocity orthogonal part of a rank-two tensor is denoted by $A^{\left\langle \mu\nu\right\rangle }=\Delta_{\alpha\beta}^{\mu\nu}A^{\alpha\beta}=\frac{1}{2}\left(\Delta_{\ \alpha}^{\mu}\Delta_{\ \beta}^{\nu}+\Delta_{\ \beta}^{\mu}\Delta_{\ \alpha}^{\nu}-\frac{2}{3}\Delta^{\mu\nu}\Delta_{\alpha\beta}\right)A^{\alpha\beta}$.

\section{First-Order Relativistic Spin Hydrodynamics}\label{chap:2}	

In this section, we briefly review the basic theoretical framework of first-order spin hydrodynamics with particle number conservation. We will introduce, in turn, the conservation equations of the system, the decomposition of physical quantities via gradient expansion, the extended thermodynamic relations, and then employ entropy current analysis (the non‑negativity condition of entropy production) to derive the constitutive relations up to first order in gradients. These constitutive relations will provide the basis for the subsequent linear mode analysis.

The equations of motion of hydrodynamics rest on very general principles: they describe the conservation of energy, momentum, and possibly conserved charge quantum numbers in the system. In spin hydrodynamics, besides the energy--momentum tensor $T^{\mu\nu}$, one must also introduce the total angular momentum tensor $J^{\lambda\mu\nu}$ \cite{Hongo:2021ona,Li:2020eon,Hattori:2019lfp,Fukushima:2020ucl,She:2021lhe,Dey:2024cwo,Biswas:2023qsw,Israel:1979wp,10.1098/rspa.1979.0005}.
\begin{align}\label{eq:1}
 \partial_{\mu}N^{\mu}=0,\quad\partial_{\mu}T^{\mu\nu}=0,\quad\partial_{\lambda}J^{\lambda\mu\nu}=0,
\end{align}
where $N^{\mu}$ denotes the particle-number current, $T^{\mu\nu}$ is the energy-momentum tensor, and $J^{\lambda\mu\nu}$ is the total angular momentum tensor. The total angular momentum tensor $J^{\lambda\mu\nu}$ can be written as 
\begin{align}\label{eq:2}
 J^{\lambda\mu\nu}
 &=L^{\lambda\mu\nu}+\Sigma^{\lambda\mu\nu} \notag \\
 &=x^{\mu}T^{\lambda\nu}-x^{\nu}T^{\lambda\mu}+\Sigma^{\lambda\mu\nu},
\end{align}
where $L^{\lambda\mu\nu}$ denotes the orbital angular momentum, while $\Sigma^{\lambda\mu\nu}$ denotes the spin angular momentum. Combining the conservation Eq. \eqref{eq:1} with the decomposition of the total angular momentum Eq. \eqref{eq:2}, one obtains the evolution equation for the spin angular momentum
\begin{align}\label{eq:3}
 \partial_{\lambda}\Sigma^{\lambda\mu\nu}=-2T^{[\mu\nu]}.
\end{align}

To construct the hydrodynamic equations, we need to perform a gradient expansion of the particle number tensor, the energy-momentum tensor, and the spin current tensor. Near local thermal equilibrium, we can decompose these tensors with respect to the four-velocity $u^\mu$ into equilibrium parts and first-order gradient corrections
\begin{gather}
 N^{\mu}=nu^{\mu}+n^{\mu}, \\
 T^{\mu\nu}=eu^{\mu}u^{\nu}-p\Delta^{\mu\nu}+2h^{(\mu}u^{\nu)}-\Pi\Delta^{\mu\nu}+\pi^{\mu\nu}+2q^{[\mu}u^{\nu]}+\phi^{\mu\nu}, \\
 \Sigma^{\lambda\mu\nu}=u^{\lambda}S^{\mu\nu}+\Sigma_{(1)}^{\lambda\mu\nu},
\end{gather}
where $nu^{\mu}$, $eu^{\mu}u^{\nu}-p\Delta^{\mu\nu}$ and $u^{\lambda}S^{\mu\nu}$ are the equilibrium parts of the respective conserved tensors, while the remaining parts of these tensors are the first-order dissipative terms from the gradient expansion. In the above expressions, $n$, $e$, $p$ and $S^{\mu\nu}$ denote the particle number density, the energy density, the isotropic pressure and the second-rank antisymmetric spin density tensor $(S^{\mu\nu}=-S^{\nu\mu})$, respectively; $n^{\mu}$, $\Pi$, $h^\mu$ and $\pi^{\mu\nu}$ represent the particle diffusion, the bulk viscous pressure, the heat flow and the shear stress tensor,respectively. As can be seen from \eqref{eq:3}, the first-order dissipative antisymmetric parts of the energy–momentum tensor, $q^\mu$ and $\phi^{\mu\nu}$, are directly related to the spin evolution. $\Sigma_{(1)}^{\lambda\mu\nu}$ is the first-order dissipative term of the spin tensor. All dissipative currents introduced above satisfy the following constraints: $n^{\mu}u_{\mu}=0$, $h^{\mu}u_{\mu}=0$, $\pi^{\mu\nu}u_{\mu}=\pi^{\mu\nu}u_{\nu}=0$, $q^{\mu}u_{\mu}=0$, $\phi^{\mu\nu}u_{\mu}=\phi^{\mu\nu}u_{\nu}=0$, $\Sigma_{(1)}^{\lambda\mu\nu}u_{\lambda}=0$. It is worth noting that the canonical energy–momentum tensor is no longer symmetric and can be further decomposed into a symmetric part and an antisymmetric part. However, this is not the only form; one can employ pseudogauge transformations to obtain different energy–momentum tensors and angular momentum tensors without altering the essential nature of the conservation laws. For instance, through such transformations one can arrive at the fully symmetric Belinfante form \cite{Fukushima:2020ucl} of the energy–momentum tensor, as well as the GLW form, among others. Nevertheless, as revealed by the pseudogauge transformation, these seemingly different frameworks differ in the description of local observables, and which one is better suited for describing experimental data from heavy‑ion collisions remains a focus of current debate.

In the present work, the~local energy density $e$, particle-number density $n$, flow velocity $u^\mu$, and~spin density $S^{\mu\nu}$ are treated as the independent hydrodynamic variables of the effective theory. Since the canonical energy-momentum tensor is generally nonsymmetric in spin hydrodynamics, we define the Landau energy frame with respect to the symmetric part of the canonical energy-momentum tensor,
\begin{align}
T_{\mathrm{S}}^{\mu\nu}\equiv T^{(\mu\nu)}.
\end{align}

The %Authors: Change the format here to first-line indent.
flow velocity and the local energy density are determined by the timelike eigenvalue problem
\begin{align}
T_{\mathrm{S}}^{\mu\nu}u_\nu=e u^\mu,\qquad u^\mu u_\mu=1 .
\end{align}
Equivalently,
\begin{align}
e=u_\mu T_{\mathrm{S}}^{\mu\nu}u_\nu,\qquad
h^\mu\equiv \Delta^\mu_{\ \alpha}T_{\rm S}^{\alpha\beta}u_\beta=0 .
\end{align}
Indeed, using the decomposition of $T_{\rm S}^{\mu\nu}$,
\begin{align}
T_{\mathrm{S}}^{\mu\nu}
=
eu^\mu u^\nu
-p\Delta^{\mu\nu}
+2h^{(\mu}u^{\nu)}
-\Pi\Delta^{\mu\nu}
+\pi^{\mu\nu},
\end{align}
one obtains
\begin{align}
T_{\rm S}^{\mu\nu}u_\nu=e u^\mu+h^\mu .
\end{align}
Thus, the Landau matching condition for $T_{\mathrm{S}}^{\mu\nu}$ is equivalent to the vanishing of the symmetric energy-diffusion current $h^\mu$. This condition fixes the energy frame and determines both $u^\mu$ and $e$, assuming that the timelike eigenvector is non-degenerate near local equilibrium. By~contrast, the~antisymmetric vector $q^\mu$ is not used to define the frame. It is associated with the antisymmetric part of the canonical energy-momentum tensor,
\begin{align}
q^\mu\equiv \Delta^\mu_{\ \alpha}T^{[\alpha\beta]}u_\beta ,
\end{align}
and is determined by its constitutive relation after the hydrodynamic variables have been~specified.

Since a dissipative fluid system is under consideration, conservation equations alone are not sufficient to mathematically close the equations of motion; constitutive relations are also required for closure. In this paper, we will employ entropy current analysis to derive the constitutive relations. Due to the inclusion of spin effects, the thermodynamic relations are extended to \cite{Hongo:2021ona,Li:2020eon,Hattori:2019lfp,Fukushima:2020ucl,She:2021lhe,Dey:2024cwo,Biswas:2023qsw}
\begin{align}
 e+p=Ts+\mu n+\omega_{\alpha\beta}S^{\alpha\beta}, \\
 {\rm d}e=T{\rm d}s+\mu{\rm d}n+\omega_{\alpha\beta}{\rm d}S^{\alpha\beta}, \\
 {\rm d}p=s{\rm d}T+n{\rm d}\mu+S^{\alpha\beta}{\rm d}\omega_{\alpha\beta},
\end{align}
where $T$, $s$, $\mu$ and $\omega_{\alpha\beta}$ denote the temperature, entropy density, chemical potential, and spin chemical potential, respectively.  In the theoretical construction of spin hydrodynamics, the order at which the spin chemical potential $\omega^{\mu \nu}$ enters the gradient expansion is a key and debated issue. On the one hand, starting from the quantum statistical density operator, one can show that in global equilibrium the spin chemical potential is proportional to the thermal vorticity $\varpi_{\mu\nu}\equiv\left(\partial_{\mu}\beta_{\nu}-\partial_{\nu}\beta_{\mu}\right)/2$, and since the thermal vorticity itself is of order $\mathcal{O}(\partial)$, it naturally follows that $\omega^{\mu \nu}\sim\mathcal{O}(\partial)$ \cite{Becattini:2012tc,Florkowski:2018ahw}. Under this counting scheme, the energy–momentum tensor generally contains an antisymmetric part, and spin and orbital angular momentum can be converted into each other. On the other hand, if one demands a fully symmetric energy–momentum tensor (e.g., the Belinfante form) \cite{Fukushima:2020ucl,She:2021lhe}, then the spin angular momentum is separately conserved. In this case, the global equilibrium condition no longer binds $\omega^{\mu \nu}$ to the thermal vorticity, allowing $\omega^{\mu \nu}\sim\mathcal{O}(1)$ as an independent zeroth-order thermodynamic variable \cite{Fukushima:2020ucl,She:2021lhe,Dey:2024cwo,Lu:2026ceo}. In the subsequent study presented in this paper, we will adopt $\omega^{\mu \nu}\sim\mathcal{O}(\partial)$ and $S^{\mu \nu}\sim\mathcal{O}(1)$.

The entropy production rate is obtained as follows\cite{Hattori:2019lfp,Fukushima:2020ucl}:
\begin{align}
 \partial_{\mu}s_{can}^{\mu}=
 &+\left(h^{\mu}-\frac{e+p}{n}n^{\mu}\right)\left[\partial_{\mu}\frac{1}{T}+\frac{1}{T}\left(u\cdot\partial\right)u_{\mu}\right] \notag \\
 &+\frac{1}{T}\pi^{\mu\nu}\partial_{\mu}u_{\nu}-\frac{1}{T}\Pi\partial_\mu u^\mu \notag \\
 &+\frac{1}{T}\phi^{\mu\nu}\left(\partial_{\mu}u_{\nu}+2\omega_{\mu\nu}\right) \notag \\
 &+\frac{q^{\mu}}{T}\left[T\partial_{\mu}\frac{1}{T}-\left(u\cdot\partial\right)u_{\mu}+4\omega_{\mu\nu}u^{\nu}\right]+\mathcal{O}\left(\partial^{3}\right),
\end{align}
where we define $u_\mu \partial^\mu \equiv \left(u\cdot\partial\right)$. To render the entropy production rate non‑negative, we require that each dissipative flux be linearly proportional to the corresponding “thermodynamic force”, i.e., we adopt the linear response approximation. From this we derive the following first‑order constitutive relations (note that the present study employs the Landau frame and $h^\mu=0$)
\begin{align}
 -\frac{e+p}{n}n^{\mu}=&\kappa\Delta^{\mu\nu}\left[\frac{1}{T}\partial_{\nu}T-\left(u\cdot\partial\right)u_{\nu}\right],\\
 \pi^{\mu\nu}=&2\eta\partial^{\langle\mu}u^{\nu\rangle},\\
 \Pi=&-\zeta\partial_{\mu}u^{\mu},\\
 q^{\mu}=&\lambda\Delta^{\mu\nu}\left[\frac{1}{T}\partial_{\nu}T+\left(u\cdot\partial\right)u_{\nu}-4\omega_{\nu\alpha}u^{\alpha}\right],\\
 \phi^{\mu\nu}=&2\gamma_{s}\Delta^{\mu\rho}\Delta^{\nu\sigma}\left(\partial_{[\rho}u_{\sigma]}+2\omega_{\rho\sigma}\right),
\end{align}
where we introduce the following transport coefficients: $\kappa$ (particle diffusion coefficient), $\eta$ (shear viscosity coefficient), $\zeta$ (bulk viscosity coefficient), $\lambda$ (coefficient related to spin–orbit coupling), and $\gamma_s$ (spin–vorticity coupling coefficient). The second law of thermodynamics demands that all transport coefficients be positive
\begin{align}\label{eq:26}
  \kappa,\eta,\zeta,\lambda,\gamma_s>0.
\end{align}

\section{Linear Mode Analysis}\label{chap:3}

In this section, we perform a linear stability and causality analysis of the first-order spin hydrodynamic equations established in Section \ref{chap:2}. We will first introduce the basic framework of linear mode analysis and the stability criteria \cite{Hiscock:1983zz,Hiscock:1985zz,Hiscock:1987zz,Daher:2022wzf,Daher:2024bah,Sarwar:2022yzs,Xie:2023gbo}, then choose a suitable static equilibrium background, introduce the plane-wave ansatz for small perturbations, and linearize the equations of motion. By solving the characteristic equation, we obtain the dispersion relation $\omega=\omega(k)$ for each perturbation mode, and analyze their behavior in the long-wavelength limit $(k\rightarrow0)$ and the short-wavelength limit $(k\rightarrow \infty)$.

Consider a fluid system in global thermodynamic equilibrium. Let all physical quantities be written as an equilibrium value plus a small perturbation: $X = X_{(0)} + \delta X$, where $X_{(0)}$ denotes the spatially uniform, time‑independent (static) equilibrium value, and $\delta X$ is a first‑order small quantity. We assume the perturbations take the form of a plane wave
\begin{align}
\delta X =\delta\tilde{X}e^{i\omega t-i\overrightarrow{\mathbf{k}} \cdot \overrightarrow{\mathbf{x}}}.
\end{align}

Under this plane wave assumption, we can get the conditions for judging stability and causality from the dispersion relation. Stability condition: All perturbation modes must decay exponentially with time, i.e., \cite{Hiscock:1983zz,Hiscock:1985zz,Hiscock:1987zz,Daher:2022wzf,Daher:2024bah,Sarwar:2022yzs,Xie:2023gbo} 
\begin{align}\label{eq:4}
\mathrm{Im}\omega\left(k\right)>0,
\end{align} 
if there exists a mode with $\operatorname{Im} \omega(k) < 0$, the perturbation grows exponentially and the system is unstable. Causality condition: The signal propagation speed must not exceed the speed of light. To this end, the following two criteria must be satisfied \cite{Hiscock:1983zz,Hiscock:1985zz,Hiscock:1987zz,Daher:2022wzf,Daher:2024bah,Sarwar:2022yzs,Xie:2023gbo,Wang:2023csj} 
\begin{align}\label{eq:5}
\underset{k\rightarrow\infty}{\lim}\left|\mathrm{Re}\frac{\omega}{k}\right|\leq1\quad \textrm{or} \quad\underset{k\rightarrow\infty}{\lim}\left|\mathrm{Re}\frac{\partial\omega}{\partial k}\right|\leq1,
\end{align}
and
\begin{align}\label{eq:6}
 \underset{k\rightarrow\infty}{\lim}\left|\frac{\omega}{k}\right| \quad\text{is bounded},
\end{align}
where the first criterion ensures that the phase velocity does not exceed the speed of light; the second criterion rules out the infinite propagation speed that occurs in non‑relativistic diffusion equations (e.g., the diffusion mode $\omega \sim i D k^2$, which satisfies the first criterion but violates the second). It should be noted that Eqs. \eqref{eq:4}-\eqref{eq:6} is a necessary but not sufficient condition for causality and stability \cite{Wang:2023csj}, yet it is adequate in most practical analyses.

For computational convenience and ease of discussion, the present work adopts a static, irrotational background
\begin{align}%\label{eq:10}
  e&=e_{(0)}+\delta e,\label{eq:7} \\
  n&=n_{(0)}+\delta n,\label{eq:8} \\
  u^\mu&=u^\mu_{(0)}+\delta u^\mu, \label{eq:9} 
\end{align}
where $u_{(0)}^{\mu}=(1,\vec{0})$, $\delta u^{\mu}=\left(0,\delta v^{i}\right)$, and $v^{i}$ the fluid three-velocity. For the spin density tensor and the spin chemical potential, under an irrotational background, it can be seen from $\varpi_{\mu\nu}\equiv\left(\partial_{\mu}\beta_{\nu}-\partial_{\nu}\beta_{\mu}\right)/2$ that the spin chemical potential vanishes. Therefore, for simplicity, one may directly assume $\omega^{\mu \nu}_{(0)}=0$ and $S^{\mu \nu}_{(0)}=0$, i.e.,
\begin{align}%\label{eq:10}
  \omega^{\mu \nu}&=0+\delta \omega^{\mu \nu},\label{eq:10}\\
  S^{\mu \nu}&=0+\delta S^{\mu \nu}.\label{eq:11}
\end{align}

In order to express other perturbations in terms of the independent perturbation variables (we choose $\delta e$, $\delta u^i$, $\delta n$, $\delta S^{0i}$, $\delta S^{ij}$), it is necessary to introduce the equation of state. One may reasonably assume that the pressure $p$ and the temperature $T$ are both functions of the energy density and the particle number density. To first order in perturbations, this can be written as
\begin{align}%\label{eq:10}
  \delta p&=c_{s}^{2}\delta e+c_{n}^{2}\delta n,\label{eq:12} \\
  \delta T&=\chi_{e}\delta e+\chi_{n}\delta n, \label{eq:13}
\end{align}
where the parameters $c_{s}^{2}$, $c_{n}^{2}$, $\chi_{e}$, and $\chi_{n}$ are treated as constants. If one considers the most general case, these parameters are all functions of the respective thermodynamic variables, such as $c_{s}^{2}=(\frac{\partial p}{\partial e})_n$ and so forth. Moreover, these parameters may not be completely independent of one another; for instance, in certain cases one can use thermodynamic relations, Maxwell relations, etc., to establish connections between different parameters.

For the spin equation of state, we adopt the simplest common form: we assume that the spin chemical potential is linearly related to the spin density, and we neglect the cross‑coupling between the energy density and the spin
\begin{align}%\label{eq:10}
  \delta\omega^{0i}=\chi_{b}\delta S^{0i},\quad \delta\omega^{ij}=\chi_{s}\delta S^{ij},\label{eq:14}
\end{align}
where $\chi_{b}$ is the proportionality parameter for the ``electric-like'' part, and $\chi_{s}$ for the ``magnetic-like'' part.

Substituting the perturbations Eqs. \eqref{eq:7}–\eqref{eq:14} into the conservation equations and constitutive relations given in Section \ref{chap:2}, and keeping terms up to first order in the perturbations, yields the perturbation equations
\begin{align}%\label{eq:10}
  0=&-\left(\kappa\beta_{(0)}h^{'}\chi_{e}\partial_{i}\partial^{i}\right)\delta\epsilon+\left(a\partial_{0}-\kappa\beta_{(0)}h^{'}\chi_{n}\partial_{i}\partial^{i}\right)\delta\mathfrak{n}+\left(\frac{1}{a}h^{'}\partial_{i}+\kappa h^{'}\partial_{i}\partial_{0}\right)\delta\vartheta^{i}, \label{eq:15} \\
  0=&\left(a\partial_{0}+\frac{1}{2}\beta_{(0)}\lambda^{'}\chi_{e}\partial^{i}\partial_{i}\right)\delta\epsilon+\left(\frac{1}{2}\beta_{(0)}\lambda^{'}\chi_{n}\partial^{i}\partial_{i}\right)\delta\mathfrak{n}+\left(\partial_{i}+\frac{1}{2}\lambda^{'}\partial_{i}\partial_{0}\right)\delta\vartheta^{i}+D_{b}\partial_{i}\delta S^{0i}, \\
  0=&-\left(\frac{1}{2}\beta_{(0)}\lambda^{'}\chi_{e}\partial_{0}\partial^{j}+ac_{s}^{2}\partial^{j}\right)\delta\epsilon-ac_{n}^{2}\partial^{j}\delta\mathfrak{n}+\left[\left(\gamma_{\parallel}-\gamma_{\perp}-\gamma^{'}\right)\partial^{j}\partial_{i}\right]\delta\vartheta^{i} \notag \\
  &+\left(\partial_{0}-\frac{1}{2}\lambda^{'}\partial_{0}\partial_{0}+\gamma_{\perp}\partial_{i}\partial^{i}+\gamma^{'}\partial_{i}\partial^{i}\right)\delta\vartheta^{j}-D_{b}\partial_{0}\delta S^{0j}+D_{s}\partial_{i}\delta S^{ij}, \\
  0=&\beta_{(0)}\lambda^{'}\chi_{e}\partial^{i}\delta\epsilon+\beta_{(0)}\lambda^{'}\chi_{n}\partial^{i}\delta\mathfrak{n}+\lambda^{'}\partial_{0}\delta\vartheta^{i}+\left(2D_{b}-\partial_{0}\right)\delta S^{0i}, \\
  0=&2\gamma^{'}\partial^{i}\delta\vartheta^{j}-2\gamma^{'}\partial^{j}\delta\vartheta^{i}+\left(2D_{s}+\partial_{0}\right)\delta S^{ij},
\end{align}
for computational simplicity, we define the following parameters
\begin{align}
\begin{array}{cccc}
\delta\epsilon \equiv \left(e_{(0)}+p_{(0)}\right)\delta e,& \delta\mathfrak{n} \equiv \left(e_{(0)}+p_{(0)}\right)\delta n, & \delta\vartheta^{i} \equiv \left(e_{(0)}+p_{(0)}\right)\delta u^{i}, 
\end{array}
\end{align}
\begin{align}
\begin{array}{cccc}
\lambda^{'} \equiv \frac{2\lambda}{e_{(0)}+p_{(0)}}, & \gamma_{\parallel} \equiv \frac{\frac{4}{3}\eta+\zeta}{e_{(0)}+p_{(0)}}, & \gamma_{\perp} \equiv \frac{\eta}{e_{(0)}+p_{(0)}}, & \gamma^{'} \equiv \frac{\gamma_{s}}{e_{(0)}+p_{(0)}},
\end{array}
\end{align}
\begin{align}
\begin{array}{cccc}
a \equiv \frac{1}{e_{(0)}+p_{(0)}}, &D_{s} \equiv 4\gamma_{s}\chi_{s}, & D_{b} \equiv 4\lambda\chi_{b},
\end{array}
\end{align}
\begin{align}
\begin{array}{cccc}
h_{(0)} \equiv \frac{n_{(0)}}{e_{(0)}+p_{(0)}}, & h^{'} \equiv \frac{h_{(0)}}{e_{(0)}+p_{(0)}}, & \beta_{(0)} \equiv \frac{1}{T_{(0)}}.
\end{array}
\end{align}

Compared with previous theoretical analyses of spin hydrodynamics, in which the particle number and the particle number conservation equation are often ignored outright, taking a non‑zero particle number into account introduces, on the one hand, an additional conservation equation, namely Eq. \eqref{eq:15}. On the other hand, since the particle number modifies the thermodynamic relations and the equation of state, it exerts distinct influences on the energy–momentum and the spin angular momentum. As a result, both the original energy–momentum conservation equations and the spin evolution equation acquire contributions from the newly included particle number--that is, perturbation terms involving the particle number appear.

Since the system is isotropic in this case, we may restrict our analysis to the $x$-direction, i.e., the plane wave takes the form
\begin{align}%\label{eq:18}
 \delta X=\delta\tilde{X}e^{i\omega t-ikx}.
\end{align}

Under the plane-wave assumption, the perturbation equations can be recast as a matrix equation
\begin{align}%\label{eq:18}
 M\delta\tilde{X}=0,
\end{align}
where
\begin{gather}
M=\begin{pmatrix}M_{1} & 0 & 0 & 0\\
0 & M_{2} & 0 & 0\\
0 & 0 & M_{2} & 0\\
0 & 0 & 0 & M_{3}
\end{pmatrix}, \\
\delta\tilde{X} \equiv \left(\delta\tilde{\mathfrak{n}},\delta\tilde{\epsilon},\delta\tilde{\vartheta}^{x},\delta\tilde{S}^{0x},\delta\tilde{\vartheta}^{y},\delta\tilde{S}^{0y},\delta\tilde{S}^{xy},\delta\tilde{\vartheta}^{z},\delta\tilde{S}^{0z},\delta\tilde{S}^{xz},\delta\tilde{S}^{yz}\right)^{\mathrm{T}},
\end{gather}
and
\begin{align}%\label{eq:18}
 M_{1}=\begin{pmatrix}i\omega a-\kappa\beta_{(0)}h^{'}\chi_{n}k^{2} & -\kappa\beta_{(0)}h^{'}\chi_{e}k^{2} & -ik\frac{1}{a}h^{'}+\kappa h^{'}k\omega & 0\\
 \frac{1}{2}\beta_{(0)}\lambda^{'}\chi_{n}k^{2} & i\omega a+\frac{1}{2}\beta_{(0)}\lambda^{'}\chi_{e}k^{2} & -ik+\frac{1}{2}\lambda^{'}\omega k & -D_{b}ik\\
 -ikac_{n}^{2} & \frac{1}{2}\beta_{(0)}\lambda^{'}\chi_{e}\omega k-ikac_{s}^{2} & \gamma_{\parallel}k^{2}+i\omega+\frac{1}{2}\lambda^{'}\omega^{2} & -i\omega D_{b}\\
 \beta_{(0)}\lambda^{'}\chi_{n}ik & \beta_{(0)}\lambda^{'}\chi_{e}ik & \lambda^{'}i\omega & 2D_{b}-i\omega
 \end{pmatrix},
\end{align}
\begin{align}%\label{eq:18}
 M_{2}=\begin{pmatrix}i\omega+\frac{1}{2}\lambda^{'}\omega^{2}+\gamma_{\perp}k^{2}+\gamma^{'}k^{2} & -i\omega D_{b} & -ikD_{s}\\
 \lambda^{'}i\omega & 2D_{b}-i\omega & 0\\
 2ik\gamma^{'} & 0 & 2D_{s}+i\omega
 \end{pmatrix},
\end{align}
\begin{align}%\label{eq:18}
 M_{3}=2D_{s}+i\omega,
\end{align}
for the equation to admit a non-trivial solution, the determinant of the matrix $M$ must vanish, a condition equivalent to
\begin{align}
 \mathrm{det}M=\mathrm{det}M_{1}\cdot\mathrm{det}M_{2}\cdot\mathrm{det}M_{2}\cdot\mathrm{det}M_{3}=0,
\end{align}
Since the dispersion relations obtained by directly solving the characteristic equations of the individual block matrices are rather involved, in this work we still first restrict ourselves to the asymptotic solutions in the limits where the wave vector $k$ tends to 0 $(k \to 0)$ and to infinity $(k \to \infty)$. As $k \to 0$, for the block matrix $M_1$, one obtains
\begin{align}\label{eq:16}
 \omega_{L,11}=\frac{-i\pm\sqrt{4D_{b}\lambda^{'}-1}}{\lambda^{'}},
\end{align}
\begin{align}\label{eq:17}
\omega_{L,12}=\pm\sqrt{\frac{x}{a}}k+i\left[\frac{1}{2}\gamma_{\parallel}+\frac{2a^{2}c_{n}^{2}h^{'}\kappa\left(x-\beta_{(0)}\chi_{e}\right)-\beta_{(0)} h^{'}\chi_{n}2ac_{n}^{2}h^{'}\kappa-\beta_{(0)} h^{'}\chi_{n}x\lambda^{'}}{4a^{2}x}\right]k^{2},
\end{align}
where, in the dispersion relations, we have introduced the parameters
\begin{align}\label{eq:18}
x\equiv ac_{s}^{2}+c_{n}^{2}h^{'}.
\end{align}
For the block matrix $M_2$, one obtains
\begin{align}\label{eq:19}
 \omega_{L,21}=\frac{-i\pm\sqrt{4D_{b}\lambda^{'}-1}}{\lambda^{'}},
\end{align}
\begin{align}\label{eq:20}
\omega_{L,22}=i\gamma_{\perp}k^{2},
\end{align}
for the block matrix $M_3$, the exact solution can be computed directly
\begin{align}\label{eq:21}
\omega_{3}=2iD_{s}.
\end{align}

The stability condition for the dispersion relations in the small wave-vector limit is given by Eq. \eqref{eq:4}
\begin{align}\label{eq:22}
\left[\frac{1}{2}\gamma_{\parallel}+\frac{2a^{2}c_{n}^{2}h^{'}\kappa\left(x-\beta_{(0)} \chi_{e}\right)-\beta_{(0)} h^{'}\chi_{n}2ac_{n}^{2}h^{'}\kappa-\beta_{(0)} h^{'}\chi_{n}x\lambda^{'}}{4a^{2}x}\right]>0,
\end{align}
\begin{align}\label{eq:23}
\lambda^{'}<0,\quad \gamma_{\perp}>0,\quad D_s>0,
\end{align}
The stability condition for the dispersion relations in the large wave-vector limit is given by Eqs. \eqref{eq:5} and \eqref{eq:6}
\begin{align}\label{eq:24}
0 \leq \sqrt{\frac{x}{a}} \leq 1 .
\end{align}

The system exhibits four non-propagating relaxation modes, given by Eqs. \eqref{eq:16}, \eqref{eq:19}, \eqref{eq:20}, and \eqref{eq:21}. Among them, the modes described by Eqs. \eqref{eq:16}, \eqref{eq:19}, and \eqref{eq:21} originate from the coupling between the spin degrees of freedom and the antisymmetric part of the energy--momentum tensor. The structure of these four non-propagating relaxation modes is identical to that of the spin relaxation modes obtained in the absence of particle-number conservation. Therefore, in the small-wave-vector limit, the spin relaxation sector remains qualitatively unchanged, indicating that the particle-number degree of freedom does not modify the intrinsic spin relaxation structure of first-order spin hydrodynamics.

On the other hand, Eq. \eqref{eq:17} corresponds to the acoustic mode of the system. Compared with the case without particle-number conservation, the propagation velocity of this mode is modified, leading to a mixing between particle-number and energy-density fluctuations. Consequently, the sound velocity depends not only on the conventional thermodynamic sound speed, $c_s$, but also on the particle-number response coefficient, $c_n$, and the background particle-number density, $h'$, as shown in Eqs. \eqref{eq:17} and \eqref{eq:18}. In particular, particle-number conservation also alters both the propagation characteristics and the stability domain of the acoustic mode. In the long-wavelength limit, the damping rate of the sound mode acquires an additional diffusive contribution, whose coefficient involves new transport and thermodynamic parameters, including $\kappa$, $\chi_n$, and $\chi_e$. As a result, unlike the case without particle-number conservation, Eq. \eqref{eq:22} shows that the stability condition is no longer determined solely by the viscous transport coefficients, but is further constrained by the particle-diffusion process. Moreover, Condition $\lambda^{'}<0$ in Eq. \eqref{eq:23} is incompatible with the entropy-positivity condition $(\lambda>0)$, implying that the inclusion of particle-number conservation does not eliminate the unstable mode present in first-order spin hydrodynamics in the small-wave-vector limit.

Next, we consider the asymptotic solution in the limit as the wave vector tends to infinity $(k\to \infty)$. For the block matrix $M_1$, one obtains
\begin{align}\label{eq:25}
\omega_{H,11}=\frac{-ia\beta_{(0)}h^{'}\kappa\chi_{n}\pm\sqrt{-\beta_{(0)}^{2}a^{2}h^{'}\kappa\chi_{n}\left(h^{'}\kappa\chi_{n}-4\lambda^{'}\chi_{e}\right)}}{2a^{2}}k^{2},
\end{align}
\begin{align}\label{eq:26}
\omega_{H,12}=i\frac{1}{L}\left(A_{1}+\frac{A_{2}}{U}+U\right),
\end{align}
\begin{align}\label{eq:27}
\omega_{H,13}=i\frac{1}{4L}\left[-4A_{1}+2\left(1+i\sqrt{3}\right)\frac{A_{2}}{U}+2\left(1-i\sqrt{3}\right)U\right],
\end{align}
\begin{align}\label{eq:28}
\omega_{H,14}=i\frac{1}{4L}\left[-4A_{1}+2\left(1-i\sqrt{3}\right)\frac{A_{2}}{U}+2\left(1+i\sqrt{3}\right)U\right],
\end{align}
the parameter $U$ introduced in the above equation is defined as follows
\begin{gather}
L\equiv6a\beta_{(0)}h^{'}\kappa\lambda^{'2}\chi_{e}\chi_{n},\\
U\equiv\left(c+\frac{1}{2}\sqrt{b+4c^{2}}\right)^{1/3},
\end{gather}
other parameters $(A_1,A_2, b, c)$ are detailed in Appendix \ref{chap:5}.
For the block matrix $M_2$, one obtains
\begin{align}\label{eq:29}
\omega_{H,21}=-2iD_{b},
\end{align}
\begin{align}\label{eq:30}
\omega_{H,22}=i\frac{2\gamma_{\perp}D_{s}}{\gamma_{\perp}+\gamma^{'}},
\end{align}
\begin{align}\label{eq:31}
\omega_{H,23}=\pm i\sqrt{\frac{2\left(\gamma_{\perp}+\gamma^{'}\right)}{\lambda^{'}}}k.
\end{align}

Modes Eqs. \eqref{eq:29}-\eqref{eq:31} are identical to the spin relaxation modes obtained without particle-number conservation. In contrast, Modes Eqs. \eqref{eq:25}-\eqref{eq:28} arise entirely from particle-number conservation and therefore constitute additional modes absent in the corresponding non-conserved case. Among them, Eqs. \eqref{eq:26}-\eqref{eq:28} represent new relaxation modes. In particular, the causality analysis of Mode Eq. \eqref{eq:25} shows that it violates the causality condition (Eq. \eqref{eq:6}), namely that the quantity $\left|\omega/k\right|$ becomes unbounded in the large-wave-vector limit. This demonstrates that the inclusion of particle-number conservation can lead to a violation of causality in the system. Since this mode is generated exclusively by particle-number conservation, it represents a genuinely non-causal mode induced by the particle-number sector.

\section{Summary and Conclusions}\label{chap:4}

In this work, we have performed a systematic analysis of the causality and stability properties of first-order relativistic spin hydrodynamics in the presence of particle-number conservation. Extending previous studies that focused primarily on charge-neutral systems, we incorporated the conserved particle-number current and its associated diffusion dynamics into the framework of canonical spin hydrodynamics and derived the complete set of linearized perturbation equations around a homogeneous equilibrium state.

The inclusion of particle-number conservation enlarges the hydrodynamic configuration space by introducing additional density and diffusion-current fluctuations. As a consequence, the collective excitation spectrum becomes richer than that of the charge-neutral theory. By analyzing the dispersion relations in both the long-wavelength and short-wavelength limits, we obtained the corresponding stability and causality constraints and clarified the role played by charge dynamics in the propagation and relaxation of hydrodynamic modes.

Our analysis shows that the spin relaxation sector is remarkably robust against the inclusion of conserved-charge dynamics. The non-propagating relaxation modes associated with spin degrees of freedom preserve the same structure as those obtained in the absence of particle-number conservation. This result indicates that the intrinsic mechanism governing spin relaxation originates from the coupling between the spin tensor and the antisymmetric part of the energy-momentum tensor and is largely insensitive to charge-density fluctuations.

In contrast, the propagating sound sector is substantially modified by particle-number conservation. The sound velocity receives additional contributions from charge-density fluctuations and thermodynamic susceptibilities, reflecting the mixing between energy-density and particle-number perturbations. Moreover, the attenuation coefficient acquires new diffusive contributions involving the particle diffusion coefficient and charge-related thermodynamic response functions. Consequently, the stability conditions are no longer determined solely by viscous transport coefficients but depend on a broader set of transport and thermodynamic parameters.

A particularly important result of this work is the emergence of new non-hydrodynamic modes induced entirely by particle-number conservation. These modes have no counterpart in the corresponding theory without conserved charges and therefore represent genuinely new collective excitations. Among them, we identified a charge-induced mode whose frequency grows without bound in the short-wavelength limit. This behavior violates the standard causality requirement that the asymptotic signal propagation speed remain finite, demonstrating that particle-number conservation can itself become a source of acausal behavior in first-order relativistic spin hydrodynamics.

Furthermore, our analysis reveals that the unstable mode already present in first-order spin hydrodynamics survives after conserved-charge dynamics is incorporated. The instability condition remains incompatible with the positivity constraints imposed by entropy production, indicating that the inclusion of particle-number conservation does not cure the fundamental pathology of the first-order theory. Instead, it introduces additional constraints and may even generate new non-causal structures.

Taken together, our results lead to two central conclusions. First, particle-number conservation does not alter the intrinsic spin relaxation structure of first-order relativistic spin hydrodynamics. Second, conserved-charge dynamics significantly modifies the propagating sector and can generate new non-hydrodynamic modes that violate causality. These findings suggest that a physically consistent description of relativistic spin fluids at finite density likely requires extensions beyond the first-order framework, analogous to the role played by Müller-Israel-Stewart theory in conventional dissipative hydrodynamics.

The present work provides a first step toward understanding the interplay between spin and conserved-charge dynamics in relativistic fluids. Future investigations may extend the analysis to second-order spin hydrodynamics, finite-spin-density backgrounds, formulations with $\omega^{\mu\nu}\sim\mathcal{O}(1)$, and alternative pseudogauge choices \cite{Lu:2026ceo}. Such studies will be essential for constructing a causal and stable theory of spin hydrodynamics applicable to finite-density matter created in heavy-ion collisions and in astrophysical environments.

\begin{acknowledgements}
	 This work is supported by the National Natural Science Foundation of China (Grants No. 12575144, and No. 11875178).
\end{acknowledgements}

\appendix
\section[\appendixname~\thesection]{} \label{chap:5}

\begin{align}
A_1 \equiv 
&-4 a^2 c_n^2 h^{'} \kappa  \lambda^{'} \chi_e+4 a^2 c_s^2 h^{'} \kappa  \lambda^{'} \chi_n-2 a^2 \gamma_{\parallel} \lambda^{'} \chi_e+2 a \beta_{(0)}  h^{'} \kappa  \lambda^{'} \chi_e \chi_n \notag \\
&+4 a^2 \gamma_{\parallel} h^{'} \kappa  \chi_n+\beta_{(0)}  h^{'} \lambda^{'2} \chi_e \chi_n,
\end{align}
\begin{align}
A_2 \equiv 
&8 a^3 \beta_(0)  h^{'} \kappa  \lambda^{'} \chi_e \chi_n (h^{'} \kappa  \chi_n (\gamma_{\parallel} (6 D_b \lambda^{'}+2)-c_s^2 \lambda^{'})+\lambda^{'} \chi_e (c_n^2 h^{'} \kappa -\gamma_{\parallel})) \notag \\
&+4 a^2 \beta_{(0)}  h^{'} \lambda^{'2} \chi_e \chi_n (h^{'} \kappa  \chi_n (-c_s^2 \lambda^{'}+2 \gamma_{\parallel}+\beta_{(0)}  \kappa  \chi_e (6 D_b \lambda^{'}+1)) \notag \\
&+\lambda^{'} \chi_e (c_n^2 h^{'} \kappa -\gamma_{\parallel}))+4 a^4 (\lambda^{'} \chi_e (2 c_n^2 h^{'} \kappa +\gamma_{\parallel})-2 h^{'} \kappa  \chi_n (c_s^2 \lambda^{'}+\gamma_{\parallel})){}^2 \notag \\
&+4 a \beta_{(0)}^2 h^{'2} \kappa  \lambda^{'3} \chi_e^2 \chi_n^2+\beta_{(0)} ^2 h^{'2} \lambda^{'4} \chi_e^2 \chi_n^2,
\end{align}
\begin{align}
b=A_2 \equiv 
&8 a^3 \beta_{(0)}  h^{'} \kappa  \lambda^{'} \chi_e \chi_n (h^{'} \kappa  \chi_n (\gamma_{\parallel} (6 D_b \lambda^{'}+2)-c_s^2 \lambda^{'})+\lambda^{'} \chi_e (c_n^2 h^{'} \kappa -\gamma_{\parallel})) \notag \\
&+4 a^2 \beta_{(0)}  h^{'} \lambda^{'2} \chi_e \chi_n (h^{'} \kappa  \chi_n (-c_s^2 \lambda^{'}+2 \gamma_{\parallel}+\beta_{(0)}  \kappa  \chi_e (6 D_b \lambda^{'}+1)) \notag \\
&+\lambda^{'} \chi_e (c_n^2 h^{'} \kappa -\gamma_{\parallel}))+4 a^4 (\lambda^{'} \chi_e (2 c_n^2 h^{'} \kappa +\gamma_{\parallel})-2 h^{'} \kappa  \chi_n (c_s^2 \lambda^{'}+\gamma_{\parallel})){}^2 \notag \\
&+4 a \beta_{(0)} ^2 h^{'2} \kappa  \lambda^{'3} \chi_e^2 \chi_n^2+\beta_{(0)} ^2 h^{'2} \lambda^{'4} \chi_e^2 \chi_n^2,
\end{align}
\begin{align}
c \equiv 
&8a^3 \beta_{(0)}^2 h^{'2} \kappa\lambda^{'3}\chi_e^2\chi_n^2(h^{'}\kappa\chi_n(-3c_s^2\lambda^{'}+\beta_{(0)}  \kappa\chi_e(9D_b\lambda^{'}+1)+\gamma_{\parallel}(9D_b\lambda^{'}+6)) \notag \\
&+3\lambda^{'}\chi_e(c_n^2h^{'}\kappa-\gamma_{\parallel}))+6a^2\beta_{(0)} ^2 h^{'2} \lambda^{'4} \chi_e^2 \chi_n^2 (h^{'} \kappa\chi_n(-c_s^2\lambda^{'}+2 \gamma_{\parallel}  \notag \\
&+2\beta_{(0)}\kappa\chi_e(3D_b\lambda^{'}+1))+\lambda^{'} \chi_e(c_n^2h^{'}\kappa-\gamma_{\parallel})) \notag \\
&+12a^4 \beta_{(0)} h^{'} \lambda^{'2} \chi_e \chi_n(2h^{'2} \kappa^2 \chi_n^2(\gamma_{\parallel}(c_s^2 \lambda^{'}+2 \beta_{(0)}\kappa\chi_e (6D_b\lambda^{'}+1)) \notag \\
&-c_s^2\lambda^{'} (c_s^2 \lambda^{'}+\beta_{(0)}\kappa\chi_e (12D_b \lambda^{'}+1))+2\gamma_{\parallel}^2) \notag \\
&+h^{'}\kappa\lambda^{'} \chi_n \chi_e(2 c_n^2 h^{'}\kappa(2 c_s^2 \lambda^{'}-\gamma_{\parallel}+\beta_{(0)}\kappa\chi_e (12D_b \lambda^{'}+1)) \notag \\
&-\gamma_{\parallel} (c_s^2\lambda^{'}+4\gamma_{\parallel}+2 \beta_{(0)}\kappa\chi_e (3D_b \lambda^{'}+1)))-(\lambda^{'2} \chi_e^2(c_n^2 h^{'} \kappa-\gamma_{\parallel}) (2 c_n^2 h^{'}\kappa+\gamma_{\parallel}))) \notag \\
&-24a^5 \beta_{(0)} h^{'} \kappa\lambda^{'} \chi_e \chi_n (\lambda^{'} \chi_e (2c_n^2 h^{'} \kappa +\gamma_{\parallel})-2h^{'} \kappa\chi_n (c_s^2 \lambda^{'}+\gamma_{\parallel})) (h^{'} \kappa \chi_n (\gamma_{\parallel} (6D_b \lambda^{'}+2) \notag \\
&-c_s^2 \lambda^{'})+\lambda^{'} \chi_e (c_n^2 h^{'} \kappa-\gamma_{\parallel}))-8a^6 (\lambda^{'} \chi_e (2c_n^2 h^{'} \kappa +\gamma_{\parallel})-2 h^{'} \kappa \chi_n (c_s^2 \lambda^{'}+\gamma_{\parallel})){}^3 \notag \\
&+6a \beta_{(0)}^3 h^{'3} \kappa  \lambda^{'5} \chi_e^3 \chi_n^3+\beta_{(0)}^3 h^{'3} \lambda^{'6} \chi_e^3 \chi_n^3,
\end{align}

\bibliography{universe_4402132_Ref}

\end{document}